\documentclass[twocolumn,prb,superscriptaddress,floatfix,showpacs]{revtex4}%

\usepackage{graphicx}
\usepackage{amsmath}
\usepackage{amsfonts}
\usepackage{amssymb}
\usepackage[ansinew]{inputenc}

\newcommand{\figwidth}{0.95\columnwidth}
\newcommand{\kF}{k_\mathrm{F}}
\newcommand{\sign}{\operatorname{sign}}
\newcommand{\step}{\operatorname{\Theta}}

\renewcommand{\Im}{\operatorname{Im}}

\begin{document}
\title{Fourier transform of the $2\kF$ Luttinger liquid density correlation function with different spin and charge velocities}
\author{Aníbal Iucci}
\affiliation{DPMC-MaNEP, University of Geneva, 24 Quai Ernest
Ansermet, CH-1211 Geneva 4, Switzerland}
\author{Gregory A. Fiete}
\affiliation{Kavli Institute for Theoretical Physics, University
of California, Santa Barbara, California 93106, USA}
\affiliation{Department of Physics, California Institute of Technology, MC 114-36, Pasadena, California 91125, USA}
\author{Thierry Giamarchi}
\affiliation{DPMC-MaNEP, University of Geneva, 24 Quai Ernest
Ansermet, CH-1211 Geneva 4, Switzerland}

\date{\today}

\begin{abstract}

We obtain a closed-form analytical expression for the zero
temperature Fourier transform of the $2\kF$ component of the
density-density correlation function in a Luttinger liquid with
different spin and charge velocities.  For frequencies near the
spin and charge singularities approximate analytical forms are
given and compared with the exact result. We find power law like
singularities leading to either divergence or cusps, depending on
the values of the Luttinger parameters and compute the
corresponding exponents. Exact integral expressions and numerical
results are given for the finite temperature case as well. We show
in particular how the temperature rounds the singularities in the
correlation function.

\end{abstract}

\pacs{71.10.Pm,71.27.+a,73.21.-b}
\maketitle


\section{Introduction}

Interacting one-dimensional systems have been proven to exhibit
exceptionally rich physics\cite{giamarchi_book_1d} such as
powerlaw decay of the correlation functions with non-universal
exponents depending on the interaction and exotic phenomena such
as spin-charge separation. Some of these phenomena have been
identified experimentally in the various realizations of one
dimensional systems such as organic conductors
\cite{schwartz_electrodynamics}, nanotubes
\cite{bockrath_luttinger_nanotubes,yao_nanotube_kink} or quantum
wires \cite{auslaender_quantumwire_tunneling_new}. The standard
paradigm describing the low energy properties of these systems is
known as Luttinger liquid (LL)
theory.\cite{haldane_bosonisation,haldane_luttinger} While the
correlation functions in LL theory are most conveniently
represented in space-time variables
\cite{giamarchi_book_1d,voit_bosonization_revue}, often
experimental results are interpreted more directly in terms of the
Fourier transform into momentum and frequency space.  Because of
the branch-cut singularity structure of the real space zero
temperature correlation functions, obtaining the Fourier space
representation is not always straightforward and requires care in
evaluating the integrals.  Perhaps this is why explicit forms of
the momentum and frequency dependent response functions have been
comparatively slow to enter the
literature.\cite{cross_spinpeierls,schulz_correlations_1d}

The Fourier transforms of the zero temperature single particle
Greens function were computed in the early
1990s\cite{voit_spectral,meden_spectral} while their finite
temperature versions followed a few years
later.\cite{schonhammer_greensFT_finte_temp,nakamura_greensFT_finite_temp}
Whereas in these references closed analytical forms for the
spinless Luttinger liquid are provided, the spinful case appears
to be much more difficult, and besides numerical results, only a
few analytical results are known\cite{kivelson_stripes_RMP}.
Moreover, even at zero temperature, an exact closed-form
expression for the $2k_F$ part of the density-density correlation
function for the spinful case appears to still be absent from the
literature\footnote{Expressions for a few special cases appear in
Ref.~[\onlinecite{orgad_spectral_functions_luttinger}].}; some
special cases of a Luther-Emery liquid with two gapped modes are
studied in Ref.~\onlinecite{orignac_appell_resonance}. In this
paper we provide an exact expression for the Fourier transform of
the density-density correlation function in the realistic case of
different spin and charge velocities and essentially arbitrary
coupling constants in the spin and charge sectors. These results
have potential implications for several experiments in one
dimensional systems. Let us mention for example Coulomb drag
between quantum
wires\cite{Fiete_coulomb_drag,Pustilnik_coulomb_drag} and
measurements of the voltage noise on a metallic gate in close
proximity to quantum wire.\cite{Fiete:prb07} In the case of strong
interactions where the spin incoherent regime can be
obtained,\cite{cheianov_incoherent_luttinger,fiete_incoherent_luttinger,Fiete:rmp07}
the finite temperature Fourier transform of the $2\kF$ density
correlations contain important information about the LL to
spin-incoherent LL crossover.

This paper is organized as follows.  In Sec.~\ref{sec:notation} we
introduce the notation and conventions we will use to describe the
spin and charge sectors (including the correlation functions) of
the LL. In Sec.~\ref{sec:zero_temp} we obtain an exact,
closed-form expression for the Fourier transform of the zero
temperature density-density correlation function in the important
case of different spin and charge velocities.  We present
approximate zero temperature results near the spin and charge
singularities in Sec.~\ref{sec:approximate} and finite temperature
results in Sec.~\ref{sec:finite_temp}. We summarize our main
points in Sec.~\ref{sec:summary}.  A few useful results and
expressions are relegated to the appendices.

\section{The spinful Tomonaga-Luttinger model}
\label{sec:notation}

The low energy properties of a 1D system of spinful fermions can
be studied with the following Hamiltonian in bosonized
form:\cite{giamarchi_book_1d}
\begin{equation}\label{eq:LuttingerHamiltonian}
H=\sum_{\nu=\rho,\sigma}\frac{v_\nu}{2\pi}\int
dx\,\left[\frac{1}{K_\nu}\left(\partial_x\phi_\nu\right)^2+K_\nu\left(\partial_x\theta_\nu\right)^2\right],
\end{equation}
where $\phi_\rho$ ($\phi_\sigma$) is a bosonic field representing
charge (spin) collective mode oscillations, $\theta_\nu$ is the
dual field satisfying
$[\phi_\nu(x),\partial_x\theta_\mu(x')]=i\pi\delta_{\nu\mu}\delta(x-x')$,
$v_\nu$ are the propagation velocities of these modes, and $K_\nu$
their stiffness constants (in this paper we take $\hbar=1$). The
Hamiltonian (\ref{eq:LuttingerHamiltonian}) becomes SU(2)
invariant for the special value $K_\sigma=1$. The fact that the
charge and spin fields commute leads to the well-known effect of
spin-charge separation, a consequence of which is the
factorization of certain correlation functions into a product of
spin and charge components when expressed as a function of space
and time. It turns out that for the line $K_\rho+K_\sigma=1$ in
parameter space this factorization also occurs in Fourier space.

In the bosonic language, the density operator has the
representation
\begin{multline}\label{eq:densityOperatorDecomposition}
\rho=\rho_0-\frac{\sqrt{2}}{\pi}\partial_x\phi_\rho
+2\rho_0\cos\left(2\kF x-\sqrt{2}\phi_\rho\right)\cos\sqrt{2}\phi_\sigma\\
+2\rho_0\cos\left(4\kF x-2\sqrt{2}\phi_\rho\right).
\end{multline}
The first term is the average density $\rho_0$, the gradient term
represents the density oscillations with zero momentum, and the
third and four terms are the $2\kF$ and $4\kF$ parts of the
density fluctuations, respectively. The decomposition
(\ref{eq:densityOperatorDecomposition}), in turn, leads to an
analogous decomposition of the Fourier transformed density-density
correlation function, $\chi(q,\omega)$:
\begin{equation}
\chi(q,\omega)=\chi_0(q,\omega)+\chi_{2\kF}(q,\omega)+\chi_{4\kF}(q,\omega),
\end{equation}
where we have neglected higher order subdominant contributions. Our main
interest is in the computation of the $2\kF$ part for arbitrary
values of the collective mode velocities $v_\nu$ and of the Luttinger parameters $K_\nu$.

In coordinate space and imaginary time, the time ordered
density-density correlation function is given by
\begin{widetext}
\begin{equation}\label{eq:density-densityCorrelationCoordinate}
\left\langle T_\tau
\rho_{2\kF}(x,\tau)\rho_{2\kF}(0,0)\right\rangle=\rho_0^2\cos(2\kF
x) \prod_{\nu=\rho,\sigma}\frac{\left(\frac{\pi\alpha}{\beta
v_\nu}\right)^{K_\nu}}{\left[\sinh\left(\frac{\pi}{\beta
v_\nu}\left(x+iv_\nu\tau-i\alpha\sign\tau\right)\right)
\sinh\left(\frac{\pi}{\beta
v_\nu}\left(x-iv_\nu\tau+i\alpha\sign\tau\right)\right)\right]^\frac{K_\nu}{2}},
\end{equation}
\end{widetext}
where $\beta$ is the inverse temperature, and $\alpha$ is a
short-distance cutoff of the order of the inverse
bandwidth.\footnote{Strictly speaking, the expression
(\ref{eq:density-densityCorrelationCoordinate}) is valid for
$(x,\tau)\gg\alpha$ and therefore, the expression of
$\chi_{2\kF}(q,\omega)$ will be valid for
$(q,\omega)\ll\alpha^{-1}$. On the other hand, neglecting $\alpha$
will only restrict the values of $K_\rho$ and $K_\sigma$ for which
our results are meaningful to the parameter region where
$K_\rho+K_\sigma<2$.} We shall follow the standard procedure to
find $\chi_{2\kF}(q,\omega)$, the Fourier transform of
Eq.(\ref{eq:density-densityCorrelationCoordinate}), and then
analytically continue the result to get the retarded function. In
spite of the simple factorized expression
(\ref{eq:density-densityCorrelationCoordinate}), a closed
analytical form of its Fourier transform has been obtained only
for the case where both spin and charge velocities are equal, or
both Luttinger parameters are equal to one.
\cite{orgad_spectral_functions_luttinger}  For the SU(2) invariant
case with arbitrary $K_\rho$, the double integral resulting from
the Fourier transform can be reduced to a single integral, but it
cannot be evaluated in closed form.  However, if we set
$i\alpha=0$ in the denominator of
\eqref{eq:density-densityCorrelationCoordinate}, the double
integral can be performed at zero temperature for arbitrary values
of $v_\nu$ and and $K_\nu$ inside the parameter regime
$K_\rho+K_\sigma<2$. Last, these results can be extended to the
region $2<K_\rho+K_\sigma<4$ where the Fourier transform bears the
same functional form than for $K_\rho+K_\sigma<2$ plus a constant
that depends on $\alpha$. In particular, the singular behavior
near $k=v_{\rho,\sigma}\omega$ is given in both regimes by
(\ref{eq:sing_behav_charge}) and (\ref{eq:sing_behav_spin})

\section{Evaluation of $\chi_{2\kF}(q,\omega)$ at zero temperature}
\label{sec:zero_temp}

The Fourier transform of
(\ref{eq:density-densityCorrelationCoordinate}) at zero
temperature can be written as
\begin{equation}\label{eq:chi_decomposition}
\chi_{2\kF}(q,i\nu)=\frac{1}{2}\left[\chi^0_{2\kF}(q+2\kF,i\nu)+\chi^0_{2\kF}(q-2\kF,i\nu)\right],
\end{equation}
with
\begin{widetext}
\begin{equation}\label{eq:fourier_transform}
\chi^0_{2\kF}(q,i\nu)=-\rho_0^2\int_{-\infty}^{\infty}dxd\tau\,
\frac{\alpha^{K_\rho}\alpha^{K_\sigma}\,e^{-i(qx-\nu\tau)}}
{\left(x^2+v_\rho^2\tau^2\right)^{\frac{K_\rho}{2}}\left(x^2+v_\sigma^2\tau^2\right)^{\frac{K_\sigma}{2}}},
\end{equation}
where we have taken $i\alpha \to0$ in the
denominators\footnote{Notice that in Eq.
(\ref{eq:fourier_transform}) we defined the Fourier transform
integral to run from $-\infty$ to $\infty$, instead of running
from $0$ to $\infty$ as one naively would do in the zero
temperature limit. In the latter case the equivalence between the
retarded and the imaginary time-ordered functions can not be
established.}. To proceed, the denominators must be combined. One
way to do this is through the introduction of another integral by
making use of Feynman's parametrization formula\cite{peskin_QFT}
\begin{equation}
\frac{1}{A^a B^b}=\frac{\Gamma(a+b)}{\Gamma(a)\Gamma(b)}
\int_0^1dw\,\frac{w^{a-1}(1-w)^{b-1}}{\left[wA+(1-w)B\right]^{a+b}},
\end{equation}
where $\Gamma$ is the gamma function, and we let
$A=x^2+v_\rho^2\tau^2$, $B=x^2+v_\sigma^2\tau^2$, $a=K_\rho/2$ and
$b=K_\sigma/2$. With this formula we find
\begin{equation}\label{eq:stepIntegral2}
\chi^0_{2\kF}(q,i\nu)=-\frac{\rho_0^2\alpha^{K_\rho+K_\sigma}
\Gamma\left(\frac{K_\rho+K_\sigma}{2}\right)}{\Gamma\left(\frac{K_\rho}{2}\right)\Gamma\left(\frac{K_\sigma}{2}\right)}
\int_{-\infty}^{\infty}dxd\tau\int_0^1dw\,
\frac{w^{\frac{K_\rho}{2}-1}(1-w)^{\frac{K_\sigma}{2}-1}e^{-i(qx-\nu\tau)}}
{\left[x^2+v_\sigma^2\tau^2+w(v_\rho^2-v_\sigma^2)\tau^2\right]^{\frac{K_\rho+K_\sigma}{2}}}.
\end{equation}
By interchanging the integration orders, one can
evaluate the integrals in $x$ and $\tau$ as
\begin{equation}\label{eq:stepIntegral3}
\chi^0_{2\kF}(q,i\nu)=-\frac{4\pi\rho_0^2\left(\frac{\alpha}{2}\right)^{K_\rho+K_\sigma}
\Gamma\left(1-\frac{K_\rho+K_\sigma}{2}\right)}{\Gamma\left(\frac{K_\rho}{2}\right)\Gamma\left(\frac{K_\sigma}{2}\right)}
\int_0^1dw\,w^{\frac{K_\rho}{2}-1}(1-w)^{\frac{K_\sigma}{2}-1}
\frac{\left[\nu^2+v_\sigma^2
q^2+\left(v_\rho^2-v_\sigma^2\right)q^2w\right]^{\frac{K_\rho+K_\sigma}{2}-1}}
{\left[v_\sigma^2+\left(v_\rho^2-v_\sigma^2\right)w\right]^{\frac{K_\rho+K_\sigma-1}{2}}},
\end{equation}
where in the last step the restriction $K_\rho+K_\sigma<2$ holds
in order to obtain a finite result. This restriction is an
artifact of letting $i\alpha\to 0$ in the denominators.  Had
$i\alpha$ been kept finite, we would have obtained a more general
result, but we would have been unable to perform the integrations
so simply. However, it is easy to extend these results to the
region $2<K_\rho+K_\sigma<4$. This can be done by noting that the
piece that diverges in the limit $\alpha\to0$ does not depend on
$k$ and $\nu$. To extract this constant we write the imaginary
exponential factor in the integrand of (\ref{eq:stepIntegral2}) as
the sum $[e^{-i(qx-\nu\tau)}-1]+1$; in the first term in square
brackets and since the divergence has been subtracted, one can
safely take the limit $\alpha\to0$, which leads to Eq.
(\ref{eq:stepIntegral3}) but now restricted to
$2<K_\rho+K_\sigma<4$. The second term just adds a cutoff
dependent constant. Finally, the integral in $w$ can be performed,
and after some algebra we find
\begin{multline}\label{eq:finalResultMatsubara}
\chi^0_{2\kF}(q,i\nu)=-\frac{4\pi\rho_0^2\left(\frac{\alpha}{2}\right)^{K_\rho+K_\sigma}
\Gamma\left(1-\frac{K_\rho+K_\sigma}{2}\right)}{\Gamma\left(\frac{K_\rho+K_\sigma}{2}\right)}
\frac{\left[\nu^2+v_\sigma^2 q^2\right]^{\frac{K_\rho+K_\sigma}{2}-1}}{v_\sigma^{K_\rho+K_\sigma-1}}\\
\times
F_1\left(\frac{K_\rho}{2},\frac{K_\rho+K_\sigma-1}{2},1-\frac{K_\rho+K_\sigma}{2},\frac{K_\rho+K_\sigma}{2};
r,1-\frac{\nu^2+v_\rho^2 q^2}{\nu^2+v_\sigma^2 q^2}\right),
\end{multline}
\end{widetext}
where $F_1$ is Appell's hypergeometric function of two
variables\cite{erdelyi_special_functions,slater_generalized_hypergeometric}
and $r=1-v_\rho^2/v_\sigma^2$. Interestingly, when
$K_\rho+K_\sigma=1$ these expressions greatly simplify. The
denominator in the integrand of equation (\ref{eq:stepIntegral3})
equals unity, and the integral reduces to a standard Gauss
hypergeometric function. Thus with $K_\rho+K_\sigma=1$ one arrives
at the simple result
\begin{multline}
\chi^0_{2\kF}(q,i\nu)=-2\pi\rho_0^2\alpha\\
\times\left(\nu^2+v_\rho^2q^2\right)^{-\frac{K_\rho}{2}}\left(\nu^2+v_\sigma^2 q^2\right)^{-\frac{K_\sigma}{2}},
\end{multline}
which shows that for this line in parameter space, the
factorization of the correlation function that represents
spin-charge separation is also obtained in Fourier space. However,
the latter factorization is non-trivial in the sense that each
factor in $\chi^0_{2\kF}(q,i\nu)$ is not the Fourier transform of
the corresponding factor in Eq.
\eqref{eq:density-densityCorrelationCoordinate}.

\section{Approximate expressions for $\chi^0_{2\kF}(q,\omega)$ near the spin and charge singularities}
\label{sec:approximate}

Physical properties of the system, i.e., measurable quantities,
can be extracted from the retarded correlation function
$\chi^0_{2\kF}(q,\omega)$. This, in turn, is related to the
Matsubara function by means of the analytic continuation
$\chi^0_{2\kF}(q,\omega)=\chi^0_{2\kF}(q,i\nu=\omega+i\delta)$.
This can be readily performed by putting a branch cut for the
power function on the negative real axis, with the convention that
it is continuous from above. Consistently, the Appell function
$F_1(\alpha,\beta,\beta',\gamma;z_1,z_2)$ has branch cuts in $z_1$
and $z_2$ along the interval $(1,\infty)$, where it is continuous
from below.\cite{olsson_analytic_continuation_appell} If we assume
that $v_\rho>v_\sigma$, and restrict ourselves to $\omega>0$, the
limit $\delta\to 0$ is straightforward, except for $v_\sigma
q<\omega<v_\rho q$, where the argument $z_2$ of $F_1$ lies over
the branch cut. As $\delta\to 0$ this line is approached from
below and therefore, we employ the continuity of the function.
Thus, we obtain the real frequency version of Eq.
\eqref{eq:finalResultMatsubara},
\begin{widetext}
\begin{multline}\label{eq:finalResultReal}
\chi^0_{2\kF}(q,\omega)=-\frac{4\pi\rho_0^2\left(\frac{\alpha}{2}\right)^{K_\rho+K_\sigma}
\Gamma\left(1-\frac{K_\rho+K_\sigma}{2}\right)}{\Gamma\left(\frac{K_\rho+K_\sigma}{2}\right)}
\frac{\left\vert \omega^2-v_\sigma^2
q^2\right\vert^{\frac{K_\rho+K_\sigma}{2}-1}
e^{-i\pi\left(\frac{K_\rho+K_\sigma}{2}-1\right)\step(\omega^2-v_\sigma^2 q^2)}}{v_\sigma^{K_\rho+K_\sigma-1}}\\
\times
F_1\left(\frac{K_\rho}{2},\frac{K_\rho+K_\sigma-1}{2},1-\frac{K_\rho+K_\sigma}{2},\frac{K_\rho+K_\sigma}{2};
r,1-\frac{\omega^2-v_\rho^2 q^2}{\omega^2-v_\sigma^2 q^2}\right),
\end{multline}
\end{widetext}
where $\step$ is the step function, a form that is well suited to
plotting.

Many times, useful information can be extracted from the behavior
near singular points of the correlation functions. Whereas in one
dimension true long-range order does not exist,\cite{giamarchi_book_1d}
for certain classes
of models, correlation functions decay algebraically for long
distances, displaying what has been called quasi long-range order.
Alternatively, they show an algebraic singular behavior when
translated into momentum space. In our result
(\ref{eq:finalResultReal}) singularities located at
$\omega=\pm v_\rho q$ and $\omega=\pm v_\sigma q$ are present, and
represents charge and spin density fluctuations respectively, a
manifestation of the spin-charge separation in a LL. In what
follows, we give the explicit behavior of
(\ref{eq:finalResultReal}) near these singular points.

The Appell function $F_1(\alpha,\beta,\beta',\gamma;z_1,z_2)$ is
defined through a double series in $z_1$ and $z_2$ within the
convergence region $|z_1|<1$ and
$|z_2|<1$.\cite{gradshteyn80_tables} It also has several singular
points, for instance $(z_1,z_2)=(0,1),\,(0,\infty)$, etc. Several
transformations permit one to obtain its analytic
continuation\cite{olsson_analytic_continuation_appell} outside
this domain by separating $F_1$ into two terms: a singular term and a term with a well defined expansion around the
singular point. This separation allows us to extract the behavior
near the charge and spin singularities:
\begin{widetext}
\begin{multline}\label{eq:chargeSingularity}
\chi^0_{2\kF}(q,\omega)\approx
-4\pi\rho_0^2v_\sigma^{1-K_\rho-K_\sigma}\left(\frac{\alpha}{2}\right)^{K_\rho+K_\sigma}
\qquad\qquad\qquad\text{for }\omega\approx \pm v_\rho q\\
\times\left(A_\rho\frac{\left\vert \omega^2-v_\rho^2
q^2\right\vert^{\frac{K_\rho}{2}+K_\sigma-1}}{\left\vert
\omega^2-v_\sigma^2 q^2\right\vert^{\frac{K_\sigma}{2}}}
e^{-i\pi\left[\left(\frac{K_\rho}{2}+K_\sigma-1\right)
\step(\omega^2-v_\rho^2
q^2)-\frac{K_\sigma}{2}\right]}+B_\rho\left\vert
\omega^2-v_\sigma^2
q^2\right\vert^{\frac{K_\rho+K_\sigma}{2}-1}e^{-i\pi\left(\frac{K_\rho+K_\sigma}{2}-1\right)}\right),
\end{multline}
\begin{multline}\label{eq:spinSingularity}
\chi^0_{2\kF}(q,\omega)\approx
-4\pi\rho_0^2v_\sigma^{1-K_\rho-K_\sigma}\left(\frac{\alpha}{2}\right)^{K_\rho+K_\sigma}
\qquad\qquad\qquad\text{for }\omega\approx \pm v_\sigma q\\
\times\left(A_\sigma\frac{\left\vert \omega^2-v_\sigma^2
q^2\right\vert^{\frac{K_\sigma}{2}+K_\rho-1}}{\left\vert
\omega^2-v_\rho^2 q^2\right\vert^{\frac{K_\rho}{2}}}
e^{-i\pi\left(\frac{K_\sigma}{2}+K_\rho-1\right)
\step(\omega^2-v_\sigma^2 q^2)}+B_\sigma\left\vert
\omega^2-v_\rho^2
q^2\right\vert^{\frac{K_\rho+K_\sigma}{2}-1}\right),
\end{multline}
\end{widetext}
where  $A_{\rho/\sigma}$ and $B_{\rho/\sigma}$ are given in Appendix~\ref{ap:coeff}.
\begin{figure}
\begin{center}
\includegraphics[width=\figwidth]{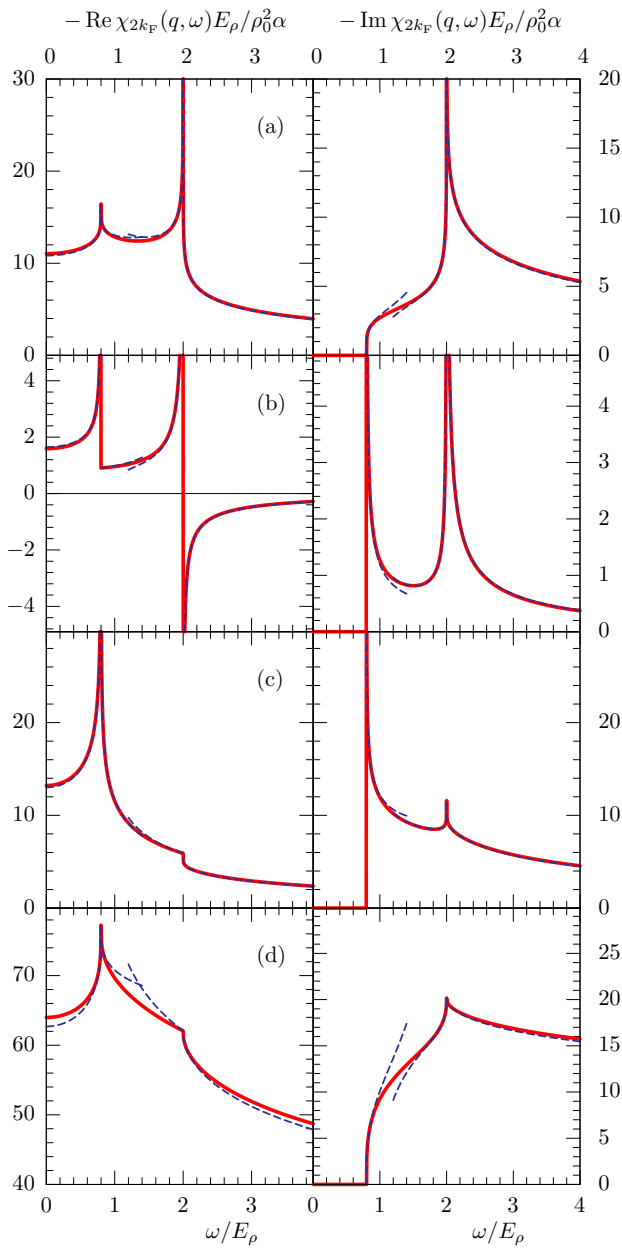}
\caption{(Color online) Real and imaginary parts of
$\chi_{2\kF}(q,\omega)$ as a function of frequency (filled red
lines) for $q=0$, $v_\sigma/v_\rho=.4$, and (a) $K_\rho=1$,
$K_\sigma=.4$, (b) $K_\rho=.4$, $K_\sigma=.2$, (c) $K_\rho=.4$,
$K_\sigma=1$ and (d) $K_\rho=.8$, $K_\sigma=1$. The last case
corresponds to the repulsive Hubbard model away from half filling.
The comparison with the approximate expressions from Eqs.
\eqref{eq:chargeSingularity} and \eqref{eq:spinSingularity} are
also shown (dashed blue lines). The charge energy is defined as
$E_\rho=\kF v_\rho$}\label{fig:freqSusceptibility}
\end{center}
\end{figure}

More schematically, the singular behavior is given by
\begin{align}
\chi^0_{2\kF}(q,\omega)&\sim|\omega^2-v_\rho^2
q^2|^{\frac{K_\rho}{2}+K_\sigma-1}+\text{const}\label{eq:sing_behav_charge}\\
\chi^0_{2\kF}(q,\omega)&\sim|\omega^2-v_\sigma^2
q^2|^{\frac{K_\sigma}{2}+K_\rho-1}+\text{const}\label{eq:sing_behav_spin}.
\end{align}
for $\omega\approx \pm v_\rho q$ and $\omega\approx \pm v_\sigma
q$ respectively. In Fig. \ref{fig:freqSusceptibility} we show
plots of the real and imaginary parts of $\chi_{2\kF}(q,\omega)$
(as defined in Eq. \eqref{eq:chi_decomposition}) as a function of
$\omega$, for a fixed value of $q=0$ and for a ratio of velocities
$v_\sigma/v_\rho=.4$. The frequency is expressed in units of the
charge energy $E_\rho=v_\rho\kF$. The singular behavior near
charge and spin singularities (\ref{eq:chargeSingularity}) and
(\ref{eq:spinSingularity}) is also shown. There will be a
divergence near the charge singularity for
$\frac{K_\rho}{2}+K_\sigma<1$ and a dip otherwise. In cases where
a divergence is present, it can be accompanied by a change of sign
when the frequency goes through the singularity ($\omega<v_\rho q$
to $\omega>v_\rho q$) as in shown Fig.
\ref{fig:freqSusceptibility}(b) (real part) [contrary to Fig.
\ref{fig:freqSusceptibility}(a)]. This is due to the imaginary
exponential factors in Eq. \eqref{eq:chargeSingularity}, and it is
easy to see that the condition to be fulfilled for a change of
sign is $\frac{K_\rho}{2}+K_\sigma-1<-\frac{1}{2}$. There are
analogous results for the spin singularity at $\omega\approx \pm
v_\sigma q$. Notice that the presence of prefactors and additional
constants in Eq. \eqref{eq:chargeSingularity} and
\eqref{eq:spinSingularity} may change dips into peaks of finite
height, as is the case of the singular behavior near the spin
singularity of the real part of $\chi_{2\kF}$ in Fig.
\ref{fig:freqSusceptibility}(a), and near the charge singularity
of the imaginary part of $\chi_{2\kF}$ in Fig.
\ref{fig:freqSusceptibility}(c) and (d).

In the figure the four possible cases are presented. In Fig.
\ref{fig:freqSusceptibility}(a) we observe a divergence in the
charge singularity in the imaginary part; in (b) both singularities are divergent; in
(c) the divergence is in the spin singularity and in (d) there are
no divergences. One especially important case is the SU(2)
symmetric line $K_\sigma=1$, where the charge singularity is
always finite, and the spin singularity can be divergent for
$K_\rho<1/2$. In particular, Eq. \eqref{eq:LuttingerHamiltonian}
represents the low energy effective Hamiltonian for the repulsive
Hubbard model away from half-filling, in which case the spin
parameter flows to the fixed point $K_\sigma^\ast=1$ and
$\frac{1}{2}<K_\rho<1$. In this situation $\chi_{2\kF}$ shows no
divergences in the spin and charge singularities, as it is
depicted in Fig. \ref{fig:freqSusceptibility}(d).

\begin{figure}
\begin{center}
\includegraphics[width=\figwidth]{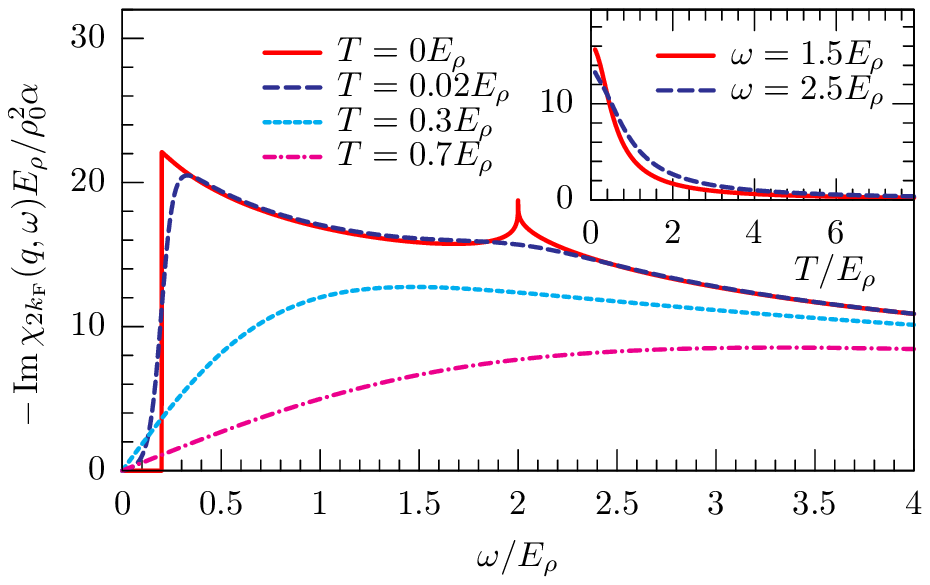}
\caption{(Color online) Imaginary parts of $\chi_{2\kF}(q,\omega)$
as a function of frequency for $q=0$, $v_\sigma/v_\rho=.1$,
$K_\rho=.5$, $K_\sigma=1$. In the inset we observe the suppression
of the correlation function under the effect of temperature (we
show its behavior for fixed values of
$\omega$).}\label{fig:spin_inc}
\end{center}
\end{figure}

\begin{figure}
\begin{center}
\includegraphics[width=\figwidth]{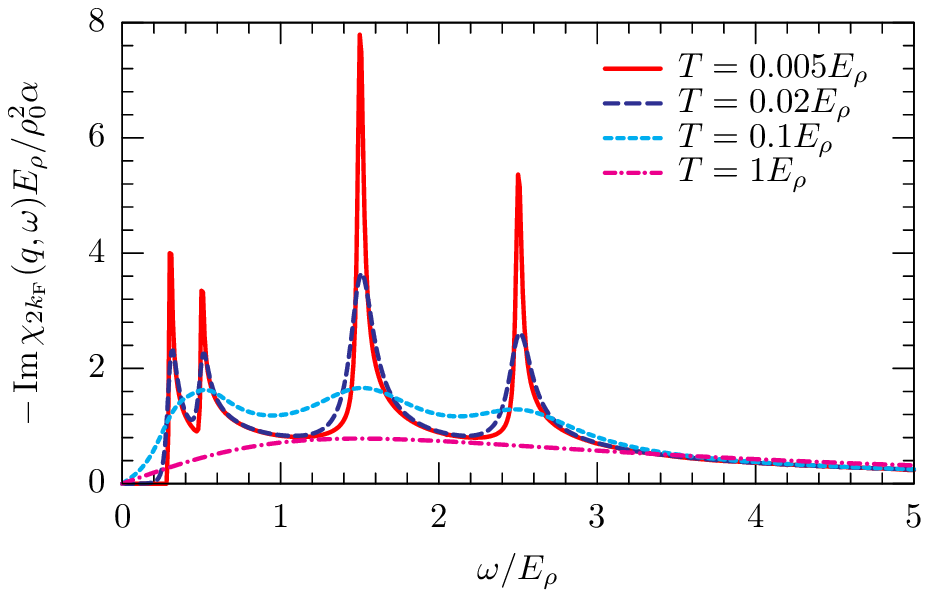}
\caption{(Color online) Imaginary parts of $\chi_{2\kF}(q,\omega)$
as a function of frequency for $q=\kF/2$, $v_\sigma/v_\rho=.2$,
$K_\rho=.4$, $K_\sigma=.2$. The four peak structure is the effect
of taking $q\neq 0$.}\label{fig:double_div}
\end{center}
\end{figure}

\section{$\chi_{2\kF}(q,\omega)$ at finite temperature}
\label{sec:finite_temp}

The finite temperature case is important as it tells how the
singularities at $\omega = \pm v_\sigma q,\pm v_\rho q$ will be
rounded with temperature.  Also, if $v_\sigma/v_\rho \ll 1$ the
spin-incoherent regime\cite{Fiete:rmp07} (defined as $v_\sigma \kF
\ll k_B T \ll v_\rho \kF$) can be approached from
temperatures below the spin energy, $v_\sigma \kF$.  As we will see
below, in this limit the $2\kF$ correlations (which may dominate
the experimental signal at zero temperature) are rapidly
suppressed with temperature. The presence or absence of $2\kF$ density
correlations at a given temperature has imporant implications for Coulomb
drag\cite{Fiete_coulomb_drag} and voltage noise\cite{Fiete:prb07} on a metallic
device in close proximity to a quantum
wire. Moreover, the temperature dependence of these quantities can reveal spin-incoherent Luttinger liquid behavior.\cite{Fiete:rmp07}

Unfortunately, we were not able to obtain a simple analytical form
for the Fourier transform of the finite temperature density
correlations \eqref{eq:density-densityCorrelationCoordinate}.
However, we were able to extract some approximate results and
study the Fourier transform numerically.

We study the temperature dependence of the Fourier transform using
the result \eqref{eq:FT_finite_T_gen},
\begin{multline}
\Im\chi^\text{ret}(q,\omega)=-\int \frac{dq'}{2\pi} \int \frac{d\epsilon'}{\pi}[n_\text{B}(\epsilon'-\omega)-n_\text{B}(\epsilon')]\\
\times
\Im\chi_\sigma^\text{ret}(q-q',\omega-\epsilon')\Im\chi_\rho^\text{ret}(q',\epsilon'),
\end{multline}
where\cite{giamarchi_book_1d}
\begin{multline}
\chi^\text{ret}_\sigma(q,\omega)=-\frac{\sin(\pi
K_\sigma/2)\rho_0\alpha^2}{v_\sigma}\left(\frac{2\pi
T\alpha}{v_\sigma}\right)^{K_\sigma-2}\\ \times
B\left(-i\frac{\omega-v_\sigma k}{4\pi
T}+\frac{K_\sigma}{4},1-\frac{K_\sigma}{2}\right)\\ \times
B\left(-i\frac{\omega+v_\sigma k}{4\pi
T}+\frac{K_\sigma}{4},1-\frac{K_\sigma}{2}\right)
\end{multline}
and $B$ is the beta function. A similar expression exists for
$\Im\chi^\text{ret}_\rho(q,\omega)$.

The behavior of the imaginary part of $\chi_{2\kF}(q,\omega)$ as a
function of $\omega$ for fixed $q$ is shown in Figs.
\ref{fig:spin_inc} (for $q=0$) and \ref{fig:double_div} (for
$q\neq0$) for two different sets of parameters. Notice that the
four peak structure of Fig. \ref{fig:double_div} is the effect of
taking $q\neq 0$. We observe that the main effect of finite
temperature is 2-fold: \emph{i}) there is a suppression of the
whole $2\kF$ correlation function, and \emph{ii}) temperature just
rounds the singularities at $\omega=\pm v_\sigma q, \pm v_\rho q$
for $|\omega -v_\sigma q|<\pi T v_\rho/v_\sigma$, while for larger
$|\omega -v_\sigma q|$ the temperature effects are minimal.

\section{Summary}
\label{sec:summary}

We have given an exact closed form expression for the zero
temperature Fourier transform of the $2\kF$ component of the
retarded density-density correlation function in a Luttinger
liquid with different velocities of spin and charge oscillations
and arbitrary stiffness constants. Additionally, we have found
approximate expressions near the collective spin and charge mode
singularities that essentially take the form of power laws, whose
exponents depend in a simple manner on $K_\rho$ and $K_\sigma$. We
also compared these approximations directly with the exact result.

We were not able to find an exact result for the finite
temperature case, but we were able to evaluate the Fourier
transform numerically and determine some approximate results.  One
important result of the analysis is that the $2\kF$ oscillations
are dramatically (exponentially) suppressed with temperature when
the spin and charge velocities are very different.  This has
implications for observable quantities that depend on the $2\kF$
density correlations, such as Coulomb drag or voltage fluctuations
on a metallic gate proximate to a quantum wire.

\acknowledgments We would like to thank M.~Cazalilla for fruitful
discussions. We gratefully acknowledge financial support from NSF
grants PHY05-51164 and DMR04-57440, the Packard Foundation, and
the Swiss National Science Foundation under MaNEP and Division II.
G.A.F. was also supported by the Lee A.~DuBridge Foundation.

\appendix
\section{Coefficients of expansions near charge and spin singularities}
\label{ap:coeff}

The coefficients $A_{\rho/\sigma}$ and $B_{\rho/\sigma}$ appearing
in \eqref{eq:chargeSingularity} and \eqref{eq:spinSingularity} are
given by
\begin{widetext}
\begin{align}
A_\rho&=\frac{\Gamma\left(1-K_\sigma-\frac{K_\rho}{2}\right)}{\Gamma\left(\frac{K_\rho}{2}\right)}%
F\left(K_\rho+K_\sigma-1,\frac{K_\rho+K_\sigma-1}{2},K_\rho+K_\sigma-1;r\right),\\
B_\rho&=\frac{\Gamma\left(1-\frac{K_\rho+K_\sigma}{2}\right)\Gamma\left(\frac{K_\rho}{2}+K_\sigma-1\right)}
{\Gamma\left(\frac{K_\sigma}{2}\right)\Gamma\left(K_\rho+K_\sigma-1\right)}
F\left(\frac{K_\rho}{2},\frac{K_\rho+K_\sigma-1}{2},K_\rho+K_\sigma-1;r\right),\\
A_\sigma&=\frac{\Gamma\left(1-K_\rho-\frac{K_\sigma}{2}\right)}{\Gamma\left(\frac{K_\sigma}{2}\right)},\\
B_\sigma&=\frac{\Gamma\left(1-\frac{K_\rho+K_\sigma}{2}\right)\Gamma\left(\frac{K_\sigma}{2}+K_\rho-1\right)}
{\Gamma\left(\frac{K_\rho}{2}\right)\Gamma\left(K_\rho+K_\sigma-1\right)}
F\left(\frac{K_\rho+K_\sigma-1}{2},\frac{K_\sigma}{2}+K_\rho-1,K_\rho+K_\sigma-1;r\right),
\end{align}
\end{widetext}
where $F$ is the Gauss Hypergeometric function (sometimes written $_2F_1$).

\section{General formula for Fourier transform of $2\kF$ response function}
\label{ap:chi_imag}

The $2\kF$ part of the density-density correlation function
\eqref{eq:density-densityCorrelationCoordinate} can be expressed
as
\begin{equation}\label{eq:chi_decomp}
\chi(x,\tau)=\chi_\sigma(x,\tau)\chi_\rho(x,\tau),
\end{equation}
where we use definition of the Fourier transform
\begin{equation}
\chi(q,i\nu)=\int_{-\infty}^{\infty} dx \int_{0}^{\beta}d\tau\,
e^{i(\nu \tau - q x)}\chi(x,\tau).
\end{equation}
to write $\chi(q,i\nu)$ as a convolution of terms
\begin{equation}
\chi(q,i\nu)=\frac{1}{\beta}\sum_{\omega_n} \int \frac{dq'}{2\pi}
\chi_\sigma(q-q',i\nu-i\omega_n)\chi_\rho(q',i\omega_n).
\end{equation}
Using the spectral representation
\begin{equation}
\chi(q,z)=-\frac{1}{\pi}\int d\epsilon \frac{\Im
\chi^\text{ret}(q,\epsilon)}{z-\epsilon},
\end{equation}
one finds
\begin{multline}
\chi(q,i\nu)=\frac{1}{\beta}\sum_{\omega_n} \int \frac{dq'}{2\pi} \int \frac{d\epsilon_1}{\pi} \int \frac{d\epsilon_2}{\pi}\\
\times \frac{\Im
\chi_\sigma^\text{ret}(q-q',\epsilon_1)}{i\nu-i\omega_n-\epsilon_1}
\frac{\Im
\chi_\rho^\text{ret}(q',\epsilon_2)}{i\omega_n-\epsilon_2},
\end{multline}
where the outer sum can be evaluated with standard complex integration:
\begin{equation}
\frac{1}{\beta}\sum_{\omega_n} \frac{1}{i\nu-i\omega_n-\epsilon_1}
\frac{1}{i\omega_n-\epsilon_2}=
-\frac{n_\text{B}(\epsilon_2)-n_\text{B}(-\epsilon_1)}{i\nu
-\epsilon_2-\epsilon_1}.
\end{equation}
Therefore, one finds
\begin{multline}
\chi(q,i\nu)=-\int \frac{dq'}{2\pi} \int \frac{d\epsilon_1}{\pi} \int \frac{d\epsilon_2}{\pi}
\frac{n_\text{B}(\epsilon_2)-n_\text{B}(-\epsilon_1)}{i\nu-\epsilon_1-\epsilon_2}\\
\times \Im \chi_\sigma^\text{ret}(q-q',\epsilon_1) \Im
\chi_\rho^\text{ret}(q',\epsilon_2),
\end{multline}
from which the analytical continuation $i\nu\to \omega +i\delta$ can readily to be done to yield
\begin{multline}\label{eq:FT_finite_T_gen}
\Im \chi(q,\omega)=-\int \frac{dq'}{2\pi} \int
\frac{d\epsilon'}{\pi}
[n_\text{B}(\epsilon'-\omega)-n_\text{B}(\epsilon')]\\
\times \Im \chi_\sigma^\text{ret}(q-q',\omega-\epsilon') \Im
\chi_\rho^\text{ret}(q',\epsilon').
\end{multline}

\end{document}